# Oxygen-vacancy-mediated Negative Differential Resistance in La and Mg co-substituted BiFeO$_3$ Thin Film

Qingqing Ke[1], Amit Kumar[2], Xiaojie Lou[1], Kaiyang Zeng[2] and John Wang[1,a]

[1] *Department of Materials Science and Engineering, National University of Singapore, Singapore 117574*

[2] *Department of Mechanical Engineering, National University of Singapore, Singapore 117576*

## ABSTRACT

The conductive characteristics of $Bi_{0.9}La_{0.1}Fe_{0.96}Mg_{0.04}O_3$ (BLFM) thin film are investigated at various temperatures and a negative differential resistance (NDR) is observed in the thin film, where a leakage current peak occurs upon application of a downward electric field above 80 °C. The origin of the NDR behavior is shown to be related to the ionic defect of oxygen vacancies ($V_O^{\bullet\bullet}$) present in the film. On the basis of analyzing the leakage mechanism and surface potential behavior, the NDR behavior can be understood by considering the competition between the polarized distribution and neutralization of $V_O^{\bullet\bullet}$.

______________________________________

a) Author to whom corresponding should be addressed. Electronic mail: msewangj@nus.edu.sg

# I. INTRODUCTION

BiFeO$_3$ (BFO), as a room temperature multiferroic material, offers great application potential in spintronics, data storage, and microelectromechanical (MEMS) devices.[1,2] Unfortunately, its wide applications are hindered by a high leakage current, which has been largely ascribed to the presence of $V_O^{\bullet\bullet}$.[3,4] Indeed, the real story is quite complicated, as $V_O^{\bullet\bullet}$ can form complex defective structure with other negative charged defects, such as $Zn'_{Fe}-V_O^{\bullet\bullet}$, which can dramatically decrease the leakage current.[5] The uncertainty triggers much interest to study the role of $V_O^{\bullet\bullet}$ in the leakage process. In addition, the interaction between $V_O^{\bullet\bullet}$ and an external electric field can give rise to rather intriguing electrical behavior. For example, the migration of $V_O^{\bullet\bullet}$ under an appropriate electric field can lead to a *p-n* junction-like behavior, and the application of an external field can be used to modulate the insulator-conductor transition by creating, erasing, and inverting the *p-n* junction.[6] Therefore, it would be of considerable interest to study the interplay between $V_O^{\bullet\bullet}$ and *dc* field in order to depict the vivid role played by $V_O^{\bullet\bullet}$ in the electric behavior of BFO thin films.

In the present work, we modulated the formation of $V_O^{\bullet\bullet}$ by doping divalent ions of Mg$^{2+}$ into the Bi$_{0.9}$La$_{0.1}$FeO$_3$ thin film (BLF) and then investigated the effect of $V_O^{\bullet\bullet}$ on its conductive behavior. A NDR behavior ($dV/dI < 0$) facilitated by $V_O^{\bullet\bullet}$ was observed in the BLFM thin films at temperatures above 80 °C. NDR effect has been observed in several ferroelectric thin films such as Pb(Zr/Ti)O$_3$ (PZT),[7] SrBi$_2$Ta$_2$O$_9$

(SBT)[8], (Pr, Mn) co-doped BFO[9] and barium titanate[10], which have been attributed to double injection,[11] trap-filling,[7] field-inhibited inter-band tunneling[12] and diffusion-limited model[13] *et al*. Notably, the previously reported NDR behaviors were likely to be inhibited by the thermal or the electric field stress, while the NDR observed in the present work is facilitated by increasing temperature, and shown to be correlated to ionic defect of $V_O^{\bullet\bullet}$ in the thin film.

## II. EXPERIMENTAL PROCEDURE

The BLFM thin film employed in the present investigation was deposited by RF magnetron sputtering on Pt-coated Si substrates at 620 °C with $SrRuO_3$ of 80 nm in thickness as the buffer layer. The detailed procedure of thin film preparation was briefly described elsewhere.[14] The resultant film exhibits a thickness of 300 nm and consists of a single perovskite phase, as verified by X-ray diffraction (XRD). Prior to electrical measurements, Au dots of 200 μm in diameter were sputtered on the films as electrodes. The leakage current was studied using a Keithley 6430 I-V system, and temperature-dependent impedance was studied using a Solartron impedance, where the measurement was conducted from 60 to 200 °C in the frequency range of $1-10^6$ Hz. To investigate the local surface charge conditions, Kelvin probe force microscopy (KPFM) was performed under ambient condition using an atomic force microscopy (MFP-3D, Asylum Research, USA) with a platinum coated silicon cantilever (radius of 15 nm with a spring constant of 2 N/m and a resonant frequency of 70 kHz).

## III. RESULTS AND DISCUSSION

Fig. 1 shows the I-V curves of the BLFM thin film at various temperatures, which can be divided into two main regions as marked I and II, respectively. In Region I, the leakage current increases almost monotonically with the increase in applied voltage. A rather anomalous leakage behavior is found in region II, where the leakage current increases with decreasing external voltage, clearly indicating an occurrence of NDR. A current peak is thus located at the crossover voltage between two regions, with a broadened shoulder extending to the lower voltage side. Inset of Fig.1 shows the I-V curve at 90 $^o$C, by sweeping the voltage as $10V \rightarrow 0V \rightarrow -10V \rightarrow 0V \rightarrow 10V$. One notes that the NDR effect in the BLFM thin film is closely related to the way of applying voltage (i.e., upward or downward sweep) and can only be observed upon application of a downward voltage (e.g., process 1 or 3 in the figure). Interestingly, the NDR effect observed here is more apparent at elevated temperatures, which is the major characteristic distinct from those in the previous investigations,[7,11-13] indicating a rather different mechanism dominating the NDR behavior in the present work.

For BFO-based thin films, the conductive behavior is largely influenced by the involved defects,[3] which can form trap states in the band gap. Therefore, the origin of the NDR effect in this work shall be considered by the defects present in the film. The frequency-dependent imaginary part of modulus ($M''$) at different temperatures is shown in Fig. 2. The loss peak is centered in the dispersion region and shifts towards the higher frequency regime with increasing temperature, indicating a thermally-activated process.[14] To identify the type of defects involved, the activation energy was calculated by fitting the loss-peak frequencies using the Arrhenius law [$f_m = f_0 \exp(-\frac{E_a}{K_B T})$], where $f_m$ is the peak frequency and $E_a$ is the activation energy. As shown in inset of Fig. 2, an excellent linear fit is shown. The estimated $E_a$=0.9 eV is comparable to the value of 0.84 eV measured for the diffusion of oxygen ions in titanate-based bulk materials.[15] This demonstrates that $V_O^{\bullet\bullet}$ are the dominant point defects in BLFM film and the formation process can be described using the Kröger–Vink notation as:

$$2MgO \rightarrow 2Mg'_{Fe} + 2O^{\times} + V_O^{\bullet\bullet} \qquad (1)$$

where $V_O^{\bullet\bullet}$ represent double-charged oxygen vacancies.

Under an applied field, $V_O^{\bullet\bullet}$ are likely to migrate towards the cathode and result in an unbalanced charge distribution throughout the film. The concentration gradient will then cause a built-in electric field which prevents the penetration of the external field. One notes that only the uncompensated $V_O^{\bullet\bullet}$ within the depletion layer contribute to the built-in field, the actual electric field ($E$) can thus be calculated by the expression [$E = E_A - \frac{2eN}{\varepsilon_0 \varepsilon}$],[16] where $N$ is the number of $V_O^{\bullet\bullet}$ aggregating within the depletion layer and $E_A$ is the applied external field. It shows that an enhancement in leakage current (shown in region II of Fig. 1) can occur due to an increase in actual electric field through decreasing the number of $V_O^{\bullet\bullet}$ across the depletion layer.

Fig. 3 shows the I-V curve with the NDR region in the logarithmic plot, based on the power law ($I \propto V^n$).[17] The plots obtained at three different temperatures are nearly parallel with each other, showing a similar trend. The plot at 100 °C can be well fitted by three linear segments with different slopes. The region with the slope of 1.3 in the lower voltage region shows a linear relationship between leakage current (*I*) and external voltage (*V*), indicating that an Ohmic conduction is the predominant factor in determining the current flow in the film, and the thermally excited electrons are believed to be the major source of current.[7] With increasing the applied voltage, more electrons, as compared to the thermally excited electrons, are injected into the thin film, then the leakage behavior would be taken over by the SCLC, obeying the modified Child' law:[18]

$$j_{SCL} = 9\varepsilon\varepsilon_0\mu\theta V^2 / 8d^3 \qquad (2)$$

where *V* is the applied voltage, *d* is film thickness, and $\theta$ is the ratio of free carrier density to the total carrier density. In Fig. 3, the derived slope of 2.01 in the region near the NDR side signifies a quadratic dependence, which is comparable to the trap-free case or the discrete trap case.[17] It therefore infers that the NDR region prior to the SCLC region involves a process of filling the traps with injected electrons.

The migration of $V_O^{\bullet\bullet}$ and their trapping of electrons under the applied electric field have been further confirmed by the KPFM studies as shown in Fig. 4. The KPFM images over a scan area of $2\times2\mu m^2$ were acquired under an ac voltage of 3 V at the lift height of 40 nm. The surface potential measured by the KPFM is a result of the compensation between the polarization bound charges and external screen charges.[19]

The dominating contribution of screen charge can be found in the case of the negative (positive) surface potential formed upon application of a negative (positive) voltage.[20] As shown in Fig. 4(a) and (b), the white and the black contrasts are observed, in comparison with the brown-colored unbiased surface, when the scan was performed with respective positive and negative bias voltage. This confirms the dominating role of screen charges originating from the injected holes and electrons from the tip upon application of an external field.[20] Fig. 4(c) shows the comparison of the surface potential profiles between the BLFM film and un-doped BLF film, indicating that the surface potential of the Mg-doped film is much higher in magnitude, regardless of the poling directions. The efficiency of the injection process is closely related to the interface condition. A lower negative surface potential observed in BLFM arises from a lower Schottky barrier height (SBH) at the interface between the tip and the film, which permits the injected electrons flow into the film easily.[20] Jeong *et al.* has showed that the SBH decreases as increasing the positive space charge density distributed at the metal/insulator interface.[21] Therefore, we argue that the lower surface potential in the BLFM film must be due to a higher concentration of $V_O^{\bullet\bullet}$ piling up at the interface, which can be neutralized by the injected electrons in the period of applying voltage.

The NDR behavior observed in this work can be explained as bellows. In the downward sweep, the large initial voltage (e.g. 10 V) can separate the defect complex ($V_O^{\bullet\bullet}$-$Mg_{Fe}^{'}$) and drive $V_O^{\bullet\bullet}$ to pile up near the interface and induce a higher built-in field, which lowers the actual field dramatically throughout the film and leads to a lower current density. Simultaneously, the dynamic injection of electrons from the cathode into the film could also occur as shown in Fig. 4. Some of the injected electrons could be localized leading to a low mobility as a result of the Anderson localization effect, or trapped by the $V_O^{\bullet\bullet}$,[6] leading to the neutralization of $V_O^{\bullet\bullet}$. Consequently, the effective number of $V_O^{\bullet\bullet}$ decreases gradually, together with a reduction in the built-in field. This results in an increase in the actual electric field, corresponding to an enhancement in leakage current as shown in region II of Fig. 1. The NDR effect dominated by the built-in field vanishes at the crossover voltage, where $V_O^{\bullet\bullet}$ are largely neutralized by the injected electrons, and the leakage current peak appears. Below the crossover voltage, the leakage current follows the variation trend of the external field. With increasing temperature, $V_O^{\bullet\bullet}$ can easily pile up in the vicinity of the cathode due to their enhanced mobility, and they should be neutralized by much more injection electrons, which gives rise to the shift of the crossover voltage towards smaller values at higher temperatures, along with the extension of the NDR region. When an opposite voltage is applied in the upward sweep (e.g., process 2 or 4 shown in inset of Fig. 1), according to Yang,[6] an inhomogeneous distribution of $V_O^{\bullet\bullet}$ could be dragged back to a homogenous state, where the $V_O^{\bullet\bullet}$ are completely compensated by $Mg'_{Fe}$. As a result, the amount of $V_O^{\bullet\bullet}$ as trapping centers is much less than that of the injected electrons, thus weakening the screen effect of $V_O^{\bullet\bullet}$ and resulting in the absence of NDR effect.

We have also considered the possibility that $V_O^{\bullet\bullet}$ accumulated near the interface may relax to an evenly distributed state due to the concentration gradient, thus resulting in a reduction of built-in field. The diffusion length of $V_O^{\bullet\bullet}$ can be estimated by $L^2 = 4Dt$,[6] in which $L$ is the diffusion length, $D$ is the diffusivity of $V_O^{\bullet\bullet}$, and $t$ is diffusion time. The diffusivity of $V_O^{\bullet\bullet}$ is in magnitude of $10^{-17}$ cm$^2$s$^{-1}$.[6] Therefore, the estimated diffusion length of $V_O^{\bullet\bullet}$ during the whole period of downward sweep (10 s) is in the order of $10^{-1}$ nm, indicating that the accumulated $V_O^{\bullet\bullet}$ are nearly frozen in the downward sweep process. The occurrence of the frozen $V_O^{\bullet\bullet}$ during the voltage sweeping process has indeed been reported in Ca-doped BFO.[6] We can conclude that the observed NDR in this work is not related to the relaxation of $V_O^{\bullet\bullet}$ back into a homogenous state.

## IV. CONCLUSIONS

To summarize, a NDR behavior has been observed in the BLFM thin film at temperatures above 80 $^\circ$C (Fig. 1) upon application of an electric field in the downward sweep. The NDR characteristics are found to be temperature-dependent and extend towards lower electric voltages side with increasing temperature. The estimated activation energy of 0.9 eV demonstrates that $V_O^{\bullet\bullet}$ are the dominant point defects involved, which are closely related to the NDR observed in this work. Through analyzing the leakage current mechanism and surface potential behavior, the NDR behavior is supposed to be in a competitive process between the polarized distribution and neutralization of $V_O^{\bullet\bullet}$ in the film.

Captions for Figures:

FIG. 1. (Color online) I-V curves of the BLFM thin film at various temperatures, by sweeping the voltage as $10V \rightarrow 0V \rightarrow -10V$, where the inset shows I-V curves measured at temperature of 90 $^oC$, by sweeping the voltage as $10V \rightarrow 0V \rightarrow -10V \rightarrow 0V \rightarrow 10V$ with the sweeping sequence denoted by the numbers. The stepwise voltage is applied with a delay time $t_d$ of 0.02 s.

FIG. 2. (Color online) Frequency dependence of $M''$. The inset shows the temperature dependence of hopping frequency obtained from $M''$ spectra for the BLFM thin film.

FIG. 3. (Color online) logI-logV curves of the BLFM thin film at various temperatures. The curve obtained at 100 $^oC$ was fitted based on power law ($I \propto V^n$).

FIG. 4. (Color online) KPFM surface potential distribution of (a) BLFM, and (b) BLF thin film. The poling area is $1.5 \times 1.5 \mu m^2$ (green rectangular) with 10 V (on the top region) and -10 V (on the bottom region) by contact mode for 1 min. (c) Surface potential profile obtained from (a) and (b) based on the line scans shown. The scale bars represent 0.5 μm.

**FIG. 1.**

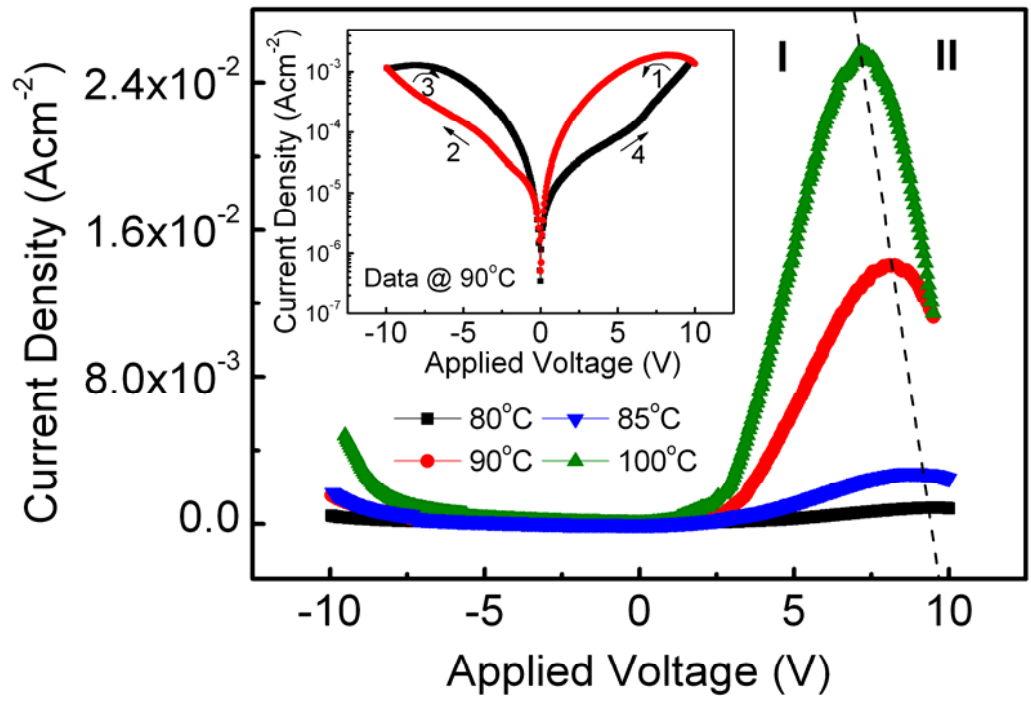

**FIG. 2.**

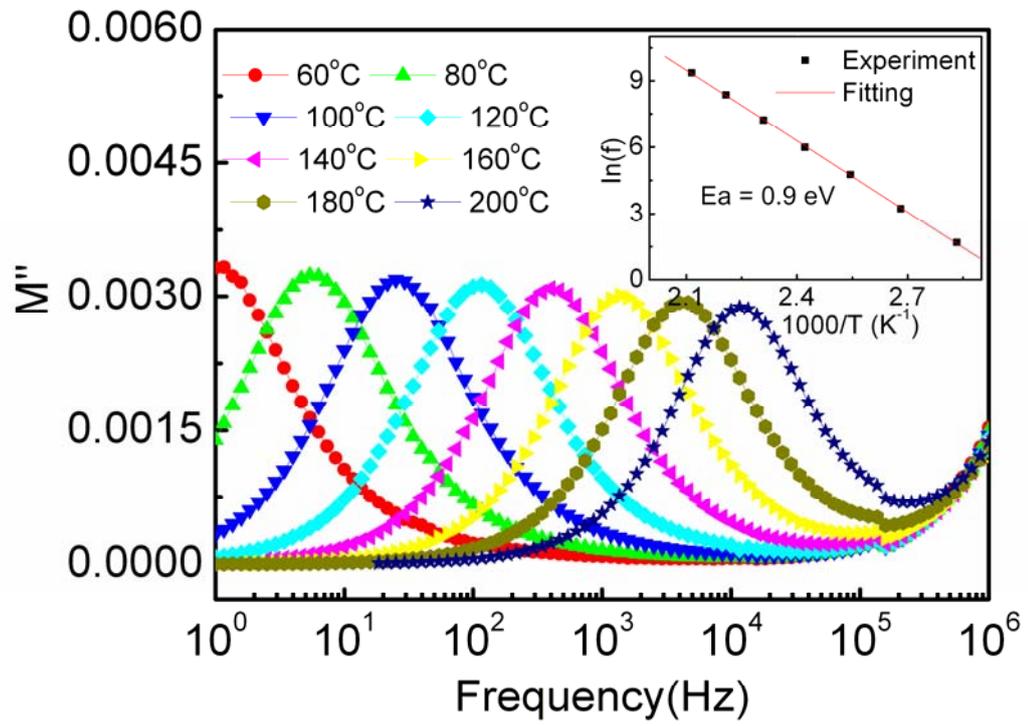

**FIG. 3.**

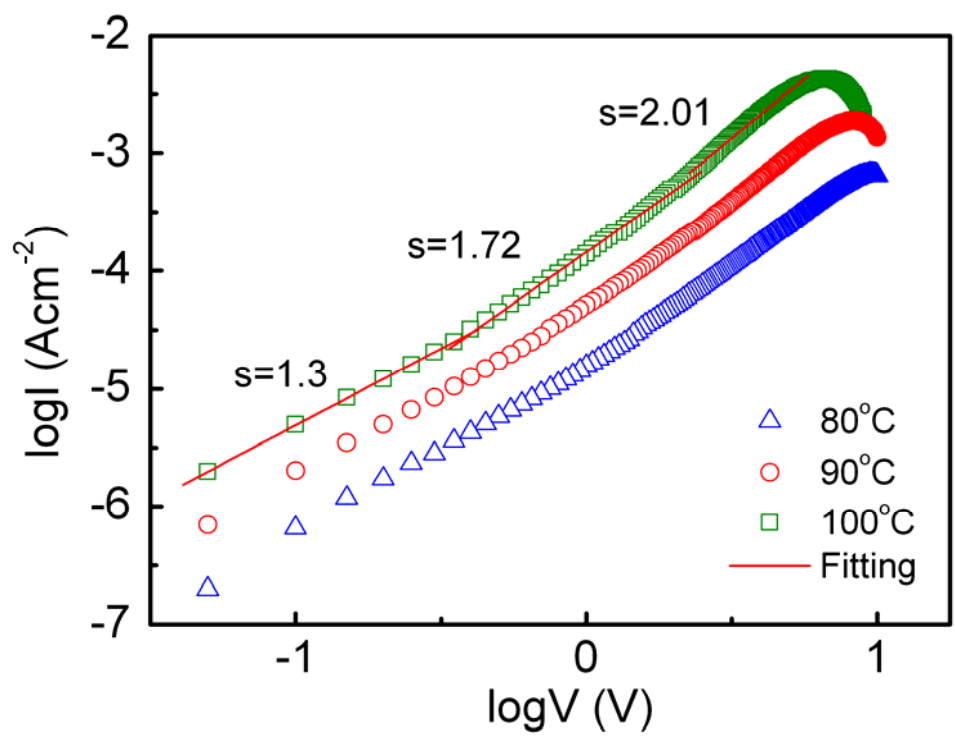

**FIG. 4.**

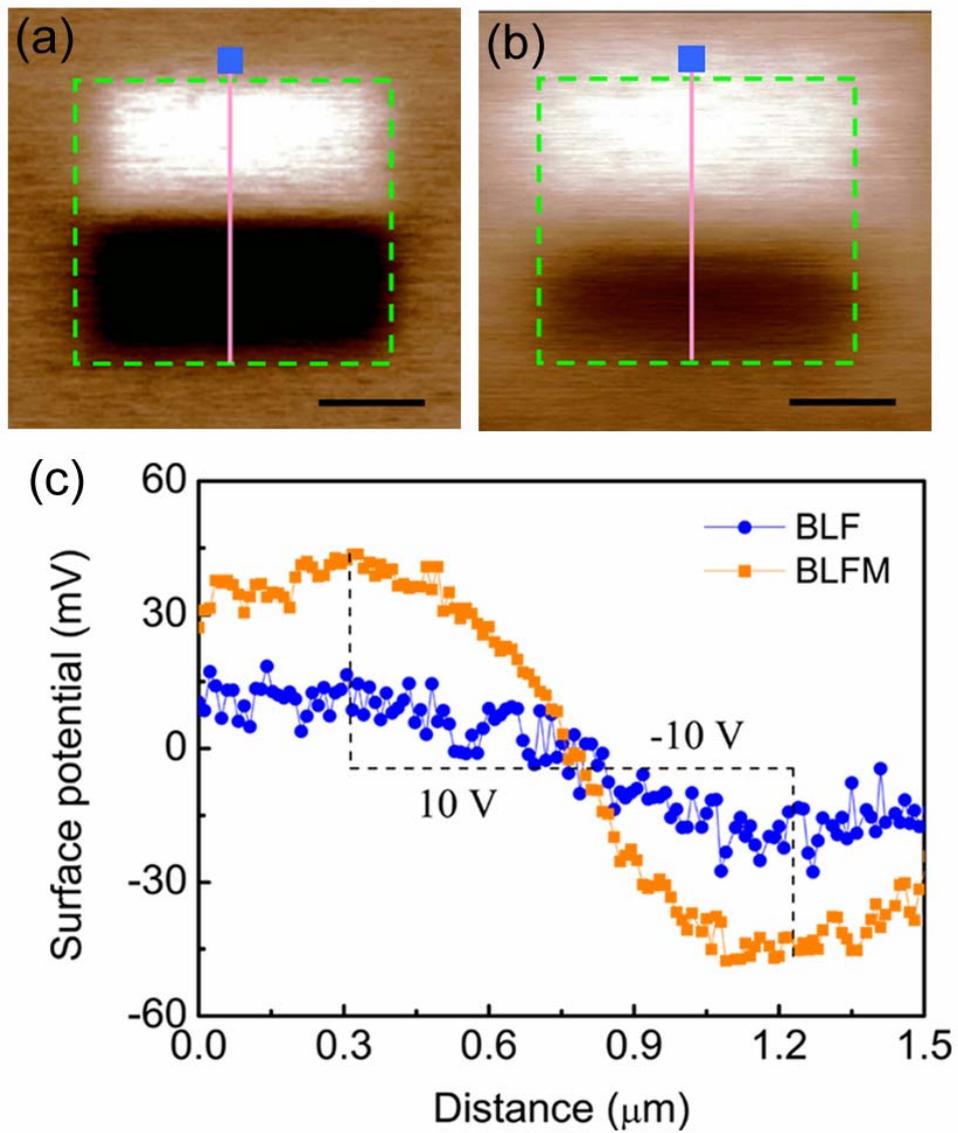